\newcommand{\CLASSINPUTtoptextmargin}{1.65cm}
\newcommand{\CLASSINPUTbottomtextmargin}{4.05cm}
\documentclass[12pt, draftclsnofoot,onecolumn]{IEEEtran}
\usepackage{graphicx} % Required for inserting images
\usepackage{color}
\usepackage{amsmath}
\usepackage{amssymb,bm}
\usepackage{latexsym}
\usepackage{cite}
\usepackage{mathbbol}

\def\mindex#1{\index{#1}}

\def\sq{\hbox{\rlap{$\sqcap$}$\sqcup$}}
\def\qed{\ifmmode\sq\else{\unskip\nobreak\hfil
\penalty50\hskip1em\null\nobreak\hfil\sq
\parfillskip=0pt\finalhyphendemerits=0\endgraf}\fi\medskip}

\long\def\defbox#1{\framebox[.9\hsize][c]{\parbox{.85\hsize}{%
\parindent=0pt
\baselineskip=12pt plus .1pt      % STYLE
\parskip=6pt plus 1.5pt minus 1pt % CHANGES
 #1}}}

\long\def\beginbox#1\endbox{\subsection*{}%
\hbox{\hspace{.05\hsize}\defbox{\medskip#1\bigskip}}%
\subsection*{}}

\def\endbox{}

\newsavebox{\junk}
\savebox{\junk}[1.6mm]{\hbox{$|\!|\!|$}}

\def\bfmath#1{{\mathchoice{\mbox{\boldmath$#1$}}%
{\mbox{\boldmath$#1$}}%
{\mbox{\boldmath$\scriptstyle#1$}}%
{\mbox{\boldmath$\scriptscriptstyle#1$}}}}

\def\bfmY{\bfmath{Y}}

\def\bfmhhaY{\bfmath{\hhaY}} %\widehat{\widehat{Y}}}}
\def\bfmhhaY{\hbox to 0pt{$\widehat{\bfmY}$\hss}\widehat{\phantom{\raise 1.25pt\hbox{$\bfmY$}}}}

\def\til={{\widetilde =}}

 \def\FRAC#1#2#3{\genfrac{}{}{}{#1}{#2}{#3}}

\def\ddtp{{\mathchoice{\FRAC{1}{d^{\hbox to 2pt{\rm\tiny +\hss}}}{dt}}%
{\FRAC{1}{d^{\hbox to 2pt{\rm\tiny +\hss}}}{dt}}%
{\FRAC{3}{d^{\hbox to 2pt{\rm\tiny +\hss}}}{dt}}%
{\FRAC{3}{d^{\hbox to 2pt{\rm\tiny +\hss}}}{dt}}}}

\def\average#1,#2,{{1\over #2} \sum_{#1}^{#2}}

\def\eye(#1){{\bf(#1)}\quad}

\def\eq#1/{(\ref{e:#1})}

\newcommand{\beqn}[1]{\notes{#1}%
\begin{eqnarray} \elabel{#1}}

\newcommand{\eeqn}{\end{eqnarray} }

\newcommand{\beq}[1]{\notes{#1}%
\begin{equation}\elabel{#1}}

\newcommand{\eeq}{\end{equation}}

\def\bdes{\begin{description}}
\def\edes{\end{description}}

\newcounter{rmnum}

\newcounter{anum}

{\end{list}}

\def\ass(#1:#2){(#1\ref{#1:#2})}

\def\ritem#1{
\item[{\sf \ass(\current_model:#1)}]
}

\newenvironment{recall-ass}[1]{%
\begin{description}
\def\current_model{#1}}{
\end{description}
}

\long\def\comment#1{}

\newfont{\bb}{msbm10 scaled 1100}

\newcommand{\av}{{\bf a}}
\newcommand{\bv}{{\bf b}}

\newcommand{\ev}{{\bf e}}

\newcommand{\gv}{{\bf g}}
\newcommand{\hv}{{\bf h}}

\newcommand{\qv}{{\bf q}}

\newcommand{\uv}{{\bf u}}
\newcommand{\wv}{{\bf w}}
\newcommand{\vv}{{\bf v}}

\newcommand{\zv}{{\bf z}}

\newcommand{\Bm}{{\bf B}}
\newcommand{\Cm}{{\bf C}}
\newcommand{\Dm}{{\bf D}}
\newcommand{\Em}{{\bf E}}
\newcommand{\Fm}{{\bf F}}
\newcommand{\Gm}{{\bf G}}

\newcommand{\Id}{{\bf I}}

\newcommand{\Pm}{{\bf P}}

\newcommand{\Rm}{{\bf R}}

\newcommand{\Tm}{{\bf T}}
\newcommand{\Um}{{\bf U}}
\newcommand{\Wm}{{\bf W}}

\newcommand{\Zm}{{\bf Z}}

\newcommand{\Cc}{{\cal C}}

\newcommand{\Nc}{{\cal N}}

\newcommand{\Pc}{{\cal P}}

\newcommand{\tauv}{\hbox{\boldmath$\tau$}}

\newcommand{\Lambdam}{\hbox{\boldmath$\Lambda$}}

\renewcommand{\arg}{{\hbox{arg}}}
\usepackage{subcaption}
\allowdisplaybreaks
\allowbreak
\title{Compressed Sensing Inspired User Acquisition for Downlink Integrated Sensing and Communication Transmissions}
\author{
	\IEEEauthorblockN{Yi Song\IEEEauthorrefmark{1}, 
     Fernando Pedraza\IEEEauthorrefmark{1},
     Shuangyang Li\IEEEauthorrefmark{1},
     Siyao Li\IEEEauthorrefmark{1}\IEEEauthorrefmark{3},
     Han Yu\IEEEauthorrefmark{2}, and 
            Giuseppe Caire\IEEEauthorrefmark{1}
	}
\IEEEauthorblockA{\IEEEauthorrefmark{1}Faculty of Electrical Engineering and Computer Science, Technical University of Berlin, Berlin, Germany}
\IEEEauthorblockA{\IEEEauthorrefmark{2}Department of Electrical Engineering, Chalmers University of Technology, Gothenburg, Sweden}
\IEEEauthorblockA{\IEEEauthorrefmark{3}Department of Electrical Engineering, University of Alaska Anchorage,  USA}
\IEEEauthorblockA{E-mail: \{yi.song, f.pedrazanieto, shuangyang.li, siyao.li, caire\}@tu-berlin.de, yuha@chalmers.se}
}

\begin{document}

\maketitle
\begin{abstract}
   This paper investigates radar-assisted user acquisition for downlink multi-user multiple-input multiple-output (MIMO) transmission using Orthogonal Frequency Division Multiplexing (OFDM) signals. 
   Specifically, we formulate a concise mathematical model for the user acquisition problem, where each user is characterized by its delay and beamspace response. Therefore, we propose a two-stage method for user acquisition, where the Multiple Signal Classification (MUSIC) algorithm is adopted for delay estimation, and then a least absolute shrinkage and selection operator (LASSO) is applied for estimating the user response in the beamspace. Furthermore, we also provide a comprehensive performance analysis of the considered problem based on the pair-wise error probability (PEP). Particularly, we show that the rank and the geometric mean of non-zero eigenvalues of the squared beamspace difference matrix determines the user acquisition performance. More importantly, we reveal that simultaneously probing multiple beams outperforms concentrating power on a specific beam direction in each time slot under the power constraint, when only limited OFDM symbols are transmitted.
   Our numerical results confirm our conclusions and also demonstrate a promising acquisition performance of the proposed two-stage method.

   %The main challenge in this process is the management of delay information, as the received signal amalgamates delays and beam space responses. 
   %Herein, we extensively compare two beam probing strategies: the beam sweep strategy and the random multi-beam strategy. %The main challenge of this problem is to handle the delay information, since the received signal is a mixture of delays and beam space responses. 
    %To mitigate the effects of delays, we propose a two-step method for user detection. Namely, we first estimate the delays based on the received signal using the Multiple Signal Classification (MUSIC) algorithm, and subsequently employ a compressed sensing method to detect users in the beam space, integrating the estimated delays. In addition, we analyze the theoretical error performance through the pairwise error probability, and provide preliminary guidelines for designing effective beam probing strategies. Numerical evaluations substantiate the efficacy of the two-stage method in user acquisition.
    %The random multi-beam strategy demonstrates superior performance over the beam sweep strategy, particularly under limited time slots. %and the random multi-beam strategy outperforms the beam sweep strategy, especially with limited resources (time slots), which is also supported by the theoretical results.

\let\thefootnote\relax\footnotetext{The work of Fernando Pedraza, Shuangyang Li, and Giuseppe Caire was supported in part by BMBF Germany in the program of ``Souver{\: a}n. Digital. Vernetzt.'' Joint Project 6G-RIC (Project IDs 16KISK030). In addition, the work of Shuangyang Li
was also supported in part by the European Union's Horizon 2020 Research and Innovation Program under MSCA Grant No. 101105732 – DDComRad. The work of Siyao Li was partially funded by the European Research Council under the ERC Advanced Grant N. 789190, CARENET. The work of Han Yu is supported in part by European Commission 101095759 Hexa-X-II. }

\end{abstract}

%\begin{keywords}
%Integrated sensing and communication, user detection, OFDM, MUSIC algorithm, compressive sensing, pairwise error probability
%\end{keywords}	
\vspace{-2mm}
 \section{Introduction}

Integrated sensing and communications (ISAC) has recently received significant attention as a key enabling technology for future wireless networks~\cite{Fan2020Joint,Weijie2021ISAC}.
Specifically, ISAC achieves both communication and radar functionalities using the same equipment, spectrum, and signals, which enjoys lower costs, higher spectral efficiency, and higher energy efficiency compared to counterparts that require dedicated transceiver designs~\cite{JointDesign,ISAC-6G,ISAC-Tradeoff,ISAC-MAC,ISAC-IT,BBP:TIT2023}.

In modern communication systems, the demand for enhanced performance and efficiency has driven the exploration of innovative techniques to optimize user acquisition processes~\cite{de2021survey}. Conventional user acquisition schemes usually 
rely on the confirmation message sent from the user side \cite{Javidi-AL}. However, thanks to the advancement of ISAC, it has been evident that user acquisition can be simplified to a target detection problem solvable by using the radar functionality, which does not require the dedicated confirmation message~\cite{liu2023integrated}. %sent by the user.

%Moreover, orthogonal frequency division multiplexing (OFDM) technology has developed continuously and has been used in the design of radar systems (see  \cite{stuber2004broadband,hwang2008ofdm,barneto2019full} and references therein). %The high bandwidth characteristic of this signal has improved the anti-jamming performance of radar system, and it has high target resolution, which has become the mainstream direction of modern radar research and development \cite{}. 
%The OFDM signal used for communication can also be applied to the present radar system and becomes an essential aspect of the integrated research of communication radar \cite{chen2021code}. 

In this paper, we focus on the radar-assisted user acquisition for downlink multiple-input multiple output (MIMO) transmissions, where a base station (BS) broadcasts orthogonal frequency division multiplexing (OFDM) signals in the cell for detecting potential users. Notice that the radar target detection based on OFDM signals is not straightforward due to the inevitable superposition among the backscattered signals from different targets and the potential frequency selectivity. Conventional methods for such a problem may rely on the matched filtering that examines all possible combinations of delay and beamspace with a dedicated resolution~\cite{Shuangyang2022Novel}, which inevitably introduces a high complexity. Hence, we are motivated to consider compressed sensing-type algorithms by noticing that the number of users is generally smaller than the size of the beamspace in a massive MIMO (mMIMO) setup.

Indeed, compressed sensing (CS) algorithms have been explored in the context of downlink frequency-division duplex (FDD) mMIMO systems for channel estimation. For instance, in~\cite{rao2014distributed}, the authors capitalize on the spatial sparsity of users' channels using discrete fourier transform (DFT) matrices and subsequently employ the joint orthogonal matching pursuit (JOMP) algorithm for estimating user responses in the angular domain. However, this approach is constrained to single carrier waveforms and lacks direct extension to the OFDM case. Another consideration is the orthogonal matching pursuit (OMP) algorithm in the multi-carrier system, as discussed in~\cite{mahdi2023FDD}, where joint estimation of angle and delay coefficients is performed. Unfortunately, this method relies on fine resolution of angle-delay grids, potentially leading to high computational complexity in practical applications.

Against this background, we propose a novel two-stage algorithm in this paper by exploiting the user sparsity in the beamspace. %by exploiting the user sparsity in the beamspace, where we assume that the users are uniformly located on all possible beam directions. 
%Specifically, 
Specifically, the proposed algorithm employs the multiple signal classification (MUSIC) algorithm for delay estimation, followed by a 
least absolute shrinkage and selection operator (LASSO) for beamspace matrix estimation. Such a scheme decouples the impact of delay from the beamspace coefficient estimation and exploits the sparsity. Furthermore, we conduct a comprehensive theoretical performance analysis utilizing pairwise error probability (PEP), and demonstrate that the rank of the squared beamspace difference matrix determines the error exponent, while the geometric mean of its non-zero eigenvalues characterizes the potential signal-to-noise ratio (SNR) improvements. Based to the PEP analysis, we reveal that probing multi-beams to different directions simultaneously generally requires less time to have a good acquisition performance compared to the conventional beam sweeping strategy, i.e., single beam at each time slot, under the same power constraint. However, as the number of time slots increases sufficiently, the beam sweeping strategy shall present a better performance. %Such a scheme decouples the impact of delay from the beamspace coefficient estimation and exploits the sparsity.
%Furthermore, we analyze the theoretical performance of the considered problem via the pair-wise error probability (PEP) framework, and demonstrate that the rank of the squared beamspace difference matrix determines the error exponent, while the geometric mean of its non-zero eigenvalues characterizes the potential signal-to-noise ratio (SNR) improvements. Based to the PEP analysis, we reveal that probing multi-beams to different directions simultaneously generally requires less time to have a good acquisition performance compared to the conventional beam sweeping strategy, i.e., single beam at each time slot, under the same power constraint. However,  when the number of time slots is sufficiently large, the beam sweeping strategy shall present a better performance. 
Numerical results confirm our conclusions from the PEP analysis and also validate the effectiveness of the proposed two-stage method.

%Numerical experiments validate the effectiveness of the proposed two-stage method and provide insights to the design of beam probing strategies.

%The main contributions of this paper can be summarized as follows, 
%\begin{itemize}
%    \item The fundamental challenge for the user detection problem lies in the unknown delay information, as the received signal is a mixture of the delay information and beamspace responses. To overcome this obstacle, we propose a novel two-stage method to exploit the delay information obtained from the received signal using the MUSIC algorithm in the first stage. In contrast, the previous work \cite{Xiaoshen2018Scalable} addressed the challenge posed by the delay information by focusing on the power of the received signal.
%    \item In this paper, we mainly focus on the two beam probing strategies, the beam sweep strategy and the random multi-beam strategy. We not only conduct some numerical experiments to compare the two strategies in terms of missed detection rate and false alarm rate as the time slot varies, but also we contribute some theoretical analysis by deriving the results based on the pairwise error probability, enhancing our understanding of these two strategies.
%\end{itemize}

% \subsection{Notations} 
\paragraph*{Notations}The superscripts $(\cdot)^{\rm{H}}$ and $(\cdot)^{\rm{T}}$ denote the Hermitian transpose and transpose of a matrix, respectively; $f^{*}(\cdot)$ and ${\rm{vec}}\left( \cdot \right)$ denote the conjugate of $f(\cdot)$ and the vectorization of a matrix; ${\rm Unif}[x, y]$ denotes the uniform distribution from $x$ to $y$; 
${\bf I}$ represents the identity matrix; 
$\mathbb{E}[\cdot]$ denotes the statistical expectation.
\vspace{-2mm}
\iffalse
\begin{table}
    \centering
    \caption{System parameters}
    \begin{tabular}{|c| c| }
    \hline
        Notations & Definitions\\
        \hline
        $N_{\rm rf}$ & number of RF chains\\
        \hline
        $N$ & number of antennas\\
        \hline
        $P$ & number of targets\\
        \hline
        $L$ & number of training time slots\\
        \hline
        $T$ & OFDM symbol duration\\
        \hline
        $M$ & number of subcarriers\\
        \hline
       % N_a & \\
    \end{tabular}
    \label{tab:my_label}
\end{table}
\fi 
\section{System Model}
\label{sec:system}
We focus on the downlink ISAC transmission utilizing OFDM signaling, functioning at a carrier frequency $f_{\rm c}$.  %which operates at a carrier frequency $f_{\rm c}$ and a bandwidth $\rm {BW}$. 
We assume that the occupied bandwidth is sufficiently smaller than $f_{\rm c}$ such that the narrow-band array response assumption holds.
Furthermore, we consider the case where the transmitter array and radar receiver array are co-located with each other at the BS, and the transmitted and received signals are perfectly separable by advanced full-duplex processing. 
% Without loss of generality, we assume that the BS equips an uniform linear arrays (ULA)  with a fully-connected hybrid digital and analog (HDA) structure, where the numbers of radio-frequency (RF) chains and antenna elements are $N_{\rm rf}$ and $M$, respectively, and the targets are located in the far-field.
We assume that the BS is equipped with a uniform linear array (ULA) with $M$ antenna elements connected to a single RF chain, and the targets are located in the far-field.

We consider the point target model, where the  $p$-th target is sufficiently characterized by its line-of-sight (LoS) path with the angle of arrival (AoA) $\phi_p$ with the marginal Doppler effect after compensated. For a time-invariant backscatter channel with $P$ separable targets{\footnote{Here, we consider the channel from the BS transmitter to the BS radar receiver, reflected by the targets.}}, its
impulse response is given by 
\begin{align}
{\bf{H}}\left( {\tau } \right) = \sum\nolimits_{p = 1}^P {{h_p}{\bf{a}}} \left( {{\phi _p}} \right){{\bf{a}}^{\rm{H}}}\left( {{\phi _p}} \right)\delta \left( {\tau  - {\tau _p}} \right), \label{channel_impulse_response}
\end{align}
where for each target $p$, $h_p$ is a complex radar channel gain including the LoS path loss, and is assumed to be a zero mean circularly-symmetric complex Gaussian variable with variance $C_{h_p} \overset{\Delta}{=}\mathbb{E}[| {{h_p}}|^2] = \frac{{{\lambda ^2}\sigma _{p, {\rm{rcs}}}}}{{{{\left( {4\pi } \right)}^3}d_p^4}}$~\cite{richards2014fundamentals}, where  
$\lambda$ is the signal wavelength, ${\sigma _{p,{\rm{rcs}}}}$ is the radar cross section (RCS) and $d_p$ is the relative distance between the $p$-th target and the BS. Moreover, $\tau_p = \frac{2 d_p}{c}$ is the round-trip delay (time of flight), where $c$ denotes the speed of light, and ${\phi_p}$ is the AoA. In~\eqref{channel_impulse_response}, ${\bf a} \left( {{\phi _p}} \right)$ is the array response vector of length $M$, whose $i$-th element is given by $[{\bf a}{(\phi _p)}]_i = e^{j\pi(i-1)\sin(\phi_p)}, 1 \le i \le {M}$.
\vspace{-1mm}
\subsection{OFDM Signaling for Target Detection}
%To detect whether a target exists in the cell, the BS first probes multiple beams towards several directions at consecutive time slots and then estimates the strength of the backscattered signals of each possible direction. If the strength is above a certain threshold, the radar receiver then declares the existence of the target. 

We focus on the signal transmission for the $l$-th OFDM symbol. Let $s_l\left( t\right)$ be the continuous-time OFDM transmitted signal without cyclic prefix (CP) at $l$-th time slot, which is written by
%\begin{align}
$s_l\left( t\right)=\sum\nolimits_{k = 1}^{N_{\rm s}} x_l\left[ k\right]p\left( t-lT\right)e^{-j2\pi \frac{k-1}{T}\left( t- lT\right)},$ %\label{OFDM_signal}
%\end{align}
where $T$ is the OFDM symbol duration, $N_{\rm s}$ is the number of subcarriers, $x_l\left[ k\right]$ is the information symbol on the $k$-th subcarrier at the $l$-th time slot satisfying $\mathbb{E}\left[\left|x_{l}[k]\right|^{2}\right] = 1$, %such that the average energy per subcarrier of an OFDM symbol is given by $E_{s}$
and $p\left( t\right)$ is the baseband shaping pulse.
As the information symbols are placed in the frequency domain, we are interested in the frequency domain channel response corresponding to~\eqref{channel_impulse_response}. Notice that~\eqref{channel_impulse_response} is time-invariant. Therefore, the frequency domain channel matrix of size $M \times M$ is also time-invariant, which can be calculated by
\begin{align}
{{\bf{\widetilde H}}_{k}} &= \int_{ - \infty }^\infty  {{\bf{H}}\left( {\tau } \right)} {e^{ - j2\pi \frac{k-1}{T}\tau }}{\rm{d}}\tau \nonumber \\
&= \sum\limits_{p = 1}^P {{h_p}{\bf{a}}} \left( {{\phi _p}} \right){{\bf{a}}^{\rm{H}}}\left( {{\phi _p}} \right){e^{ - j2\pi \frac{k-1}{T}{\tau _p}}}, ~\forall k \in [N_{\rm s}],\label{chanel_freq}
\end{align}
where $k$ is the subcarrier index.
Thus, by transmitting OFDM signal $s_l(t)$ %~\eqref{OFDM_signal} 
over the channel characterized by~\eqref{channel_impulse_response} with a sufficiently long CP, and considering the same beamforming vector $\vv_{l}$ is applied at the transmitter and the receiver, we can derive the frequency domain channel observations after the CP removal and matched-filtering as 
\begin{align}
{y_l}\left[ k \right] = {\bf{v}}_l^{\rm{H}}{{{\bf{\widetilde H}}}_{k}}{{\bf{v}}_l}{x_l}\left[ k \right] + {n_l}\left[ k \right], \label{y_l_k}
\end{align}
where $y_l\left[ k\right]$ and $n_l\left[ k\right]$ are the received symbol and the noise sample on the $k$-th subcarrier at the $l$-th time slot, respectively. 
% Particularly, ${\bf v}_l$ is the weighted sum of the columns of the beamforming matrix ${\bf{F}} = \left[ {{{\bf{f}}_1},{{\bf{f}}_2},\cdots,{{\bf{f}}_{{N_{\rm rf}}}}} \right]$, i.e., 
We assume that the BS makes use of a beamforming codebook basis ${\bf{F}} = [{\bf{f}}_{1}, {\bf{f}}_{2}, \dots, {\bf{f}}_{N_{\rm b}}]$ of cardinality $N_{\rm b}$, where $N_{\rm b} < M$. The adopted beamforming vectors are synthesized as a linear combination of the codebook entries, i.e.,
\begin{align}\label{multibeam}
{{\bf{v}}_l} = \sum\nolimits_{i = 1}^{{N_{\rm b}}} {{w_{l,i}}{{\bf{f}}_i}},
\end{align}
where $0 \le w_{l,i} \le 1$ is the weight for the $i$-th beamformer ${{\bf{f}}_i}$ at the $l$-th time slot.
Focusing on codebooks with approximately pairwise orthonormal columns, i.e.,  $\,\Fm^{\rm H}\Fm \approx \Id$, we further constraint the weights to satisfy $\sum\nolimits_{i = 1}^{{N_{\rm b}}} {{{\left| {{w_{l,i}}} \right|}^2}}  = 1$ such that $\|{\bf{v}}_{l}\|^2 = 1$. Specifically, we consider codebook vectors with approximately constant gain within their beamwidths and low gain elsewhere, designed as described in \cite[App. A]{SaeidBeamforming}. Furthermore, we construct the codebook such that each angle is covered by a single beam, i.e., $|{\bf f}_{i}^{\rm H}{\bf a}(\phi)| \gg 1$ implies $|{\bf f}_{j}^{\rm H}{\bf a}(\phi)| \approx 0$ for $i \ne j$.
In such a case, the noise sample ${z_l}\left[ k \right]$ is a zero mean complex Gaussian variable, i.e., ${\mathbb E}\left[ {{n_l}\left[ k \right]{{ {{n_l^*}\left[ k \right]} }}} \right] = {N_0}{\bf{v}}_l^{\rm{H}}{{\bf{v}}_l}=N_0$, where  $N_0$ is the one-sided power spectral density of the underlying additive white Gaussian noise (AWGN) process.  

By substituting~\eqref{multibeam} into~\eqref{y_l_k}, the received signal is given by
\begin{align}
{y_l}\left[ k \right] &=  {\bf{v}}_l^{\rm{H}}{{{\bf{\widetilde H}}_k}}{{\bf{v}}_l} {x_l}\left[ k \right] + {n_l}\left[ k \right] \nonumber \\
&= \sum\limits_{i = 1}^{{N_{\rm b}}} {\sum\limits_{j = 1}^{{N_{\rm b}}} {{w_{l,i}}} } w_{l,j}^*{\bf{f}}_j^{\rm{H}}{{{\bf{\widetilde H}}}_k}{{\bf{f}}_i}{x_l}\left[ k \right]  + {n_l}\left[ k \right] .\label{y_l_k_simplified}
\end{align}
%\nbl{SL: Should not the summation in \eqref{y_l_k_simplified} (i.e., from $i=1$ to $i= N_{\rm rf}$) match with \eqref{multibeam} (i.e., from $i=1$ to $i= N$)? }
For the $k$-th subcarrier, note that $|{\bf f}_{j}^{\rm H} {\bf a}(\phi_{p}){\bf a}^{\rm H}(\phi_{p}){\bf f}_{i}| \approx 0$ for $i \ne j$ and all $\phi_{p}$ holds due to the codebook. Hence, based on the definition of ${\bf{\widetilde H} }_k$ in \eqref{chanel_freq}, \eqref{y_l_k_simplified} 
can be further simplified as 
%\begin{align}\label{eq:y_l_k}
    ${y_l}\left[ k \right] = \sum_{i=1}^{N_{\rm b}} |w_{l, i}|^2 {\bf{f}}_i^{\rm H}  {\bf{\widetilde H}}_k {\bf{f}}_i x_l[k] + n_l[k].$
%\end{align}
Note that ${x_l}\left[ k \right]$ as the pilot information is known at the radar receiver, so we are able to define the received signal as ${r_l}\left[ k \right] \overset{\Delta}{=} {y_l}\left[ k \right]/{x_l}\left[ k \right]$ and the effective noise as ${z_l}\left[ k \right] \overset{\Delta}{=} {n_l}\left[ k \right]/{x_l}\left[ k \right]$, and rewrite ${y_l}\left[ k \right]$ %
%~\eqref{eq:y_l_k} 
based on the channel formulation in \eqref{chanel_freq} as 
\begin{align}\label{r_l_k_on-grid}
&{r_l}\left[ k \right] = \sum_{i=1}^{N_{\rm b}} |w_{l, i}|^2 {\bf{f}}_i^{\rm H}  {\bf{\widetilde H}}_k {\bf{f}}_i  \!+\! z_l[k] \nonumber \\
&= \sum_{i=1}^{N_{\rm b}} \sum_{p=1}^P |w_{l, i}|^2  h_p {e^{ - j2\pi \frac{k-1}{T}{\tau _p}}} {\bf{f}}_i^{\rm H} \av(\phi_p) \av(\phi_p)^{\rm H} {\bf{f}}_i \!+\! z_l[k]
\end{align}
where ${z_l}\left[ k \right]$ has the same variance of ${n_l}\left[ k \right]$ for energy normalized ${x_l}\left[ k \right]$.
Furthermore, by stacking ${r_l}\left[ k \right]$ for all $k \in [N_{\rm s}]$ into a vector, the observations at the $l$-th time slot is given by 
\begin{align}\label{r_l_on-grid}
{{\bf{r}}_l} 
% &\overset{\Delta}{=}\begin{bmatrix}
% r_l[1]\\
% r_l[2] \\
% \cdot \cdot \cdot\\
% r_l[N_{\rm s}]
% \end{bmatrix} \nonumber \\
\!=&\! \begin{bmatrix}
\sum_{i=1}^{N_{\rm b}} \sum_{p=1}^P |w_{l, i}|^2  h_p {\bf{f}}_i^{\rm H} \av(\phi_p) \av(\phi_p)^{\rm H} {\bf{f}}_i\\
\sum_{i=1}^{N_{\rm b}} \sum_{p=1}^P |w_{l, i}|^2  h_p {e^{ - j2\pi \frac{1}{T}{\tau_p}}} {\bf{f}}_i^{\rm H} \av(\phi_p) \av(\phi_p)^{\rm H} {\bf{f}}_i \\
\vdots\\
\sum_{i=1}^{N_{\rm b}} \sum_{p=1}^P |w_{l, i}|^2  h_p {e^{ - j2\pi \frac{N_{\rm s}-1}{T}{\tau_p}}} {\bf{f}}_i^{\rm H} \av(\phi_p) \av(\phi_p)^{\rm H} {\bf{f}}_i
\end{bmatrix}\nonumber \\
& + 
\left[z_l\left[1\right],z_l\left[2\right],...,z_l\left[N_s\right]\right]^{\rm T} 
\overset{\Delta}{=} \Tm \wv_l + \zv_l,
\end{align}
where for notational simplicity, we let 
\begin{align}
{{\bf{w}}_l} = {\left[ {{{\left| {{w_{l,1}}} \right|}^2},{{\left| {{w_{l,2}}} \right|}^2},\cdots,{{\left| {{w_{l, N_{\rm b}}}} \right|}^2}} \right]^{\rm{T}}} \label{def:bfw_l}
\end{align}
as the \textit{beam weight vector} for the $l$-th time slot, and let the effective noise vector as $\zv_l = [z_l[1], \cdots, z_l[N_{\rm s}]]^{\rm T}$. More importantly, we define 
${\bf{T}}\in \mathbb{C}^{N_{\rm s}\times N_{\rm b}}$ as the \textit{radar response matrix} and it can be decomposed as 
%\begin{align}\label{eq:q_b_g}
    $\Tm = \sum_{p=1}^P  h_p \bv(\tau_p) \qv_p^{\rm T}, $
%\end{align}
where $\bv(\tau_p)$ is the \textit{delay steering vector} of length-$N_{\rm s}$ represented by ${\bf{b}}\left( {{\tau _p}} \right) \buildrel \Delta \over = {\left[ {1,{e^{ - j2\pi \frac{1}{T}{\tau _p}}},\cdots,{e^{ - j2\pi \frac{{N_{\rm s} - 1}}{T}{\tau _p}}}} \right]^{\rm{T}}},$
%\begin{align*}
   %{\bf{b}}\left( {{\tau _p}} \right) \buildrel \Delta \over = {\left[ {1,{e^{ - j2\pi \frac{1}{T}{\tau _p}}},\cdots,{e^{ - j2\pi \frac{{N_{\rm s} - 1}}{T}{\tau _p}}}} \right]^{\rm{T}}}, 
%\end{align*}
 and $\qv_p$ denotes the \textit{target beamspace vector} of length-$N_{\rm b}$ characterizing the reflectivity of the $p$-th target with respect to each possible beam direction, given as
% \begin{align*}
$ {{\bf{q}}_p} = [{{\bf{f}}_1^{\rm{H}}{\bf{a}}\left( {{\phi _p}} \right){{\bf{a}}^{\rm{H}}}\left( {{\phi _p}} \right){{\bf{f}}_1}}, \cdots, {{\bf{f}}_{N_{\rm b}}^{\rm{H}}{\bf{a}}\left( {{\phi _p}} \right){{\bf{a}}^{\rm{H}}}\left( {{\phi_p}} \right){{\bf{f}}_{N_{\rm b}}}}]^{\rm T}.$
%\end{align*}
Since we could not decouple the effect of $h_p$ for the user detection problem, we also define $\gv_p \overset{\Delta}{=} h_p \qv_p$ and our goal is to estimate $\gv_p$, which contains the information of the user location in the beamspace, based on the observations collected within the $L$ time slots.
If we stack the delay steering vectors for all delays as $\Bm \overset{\Delta}{=} [\bv(\tau_1), \cdots, \bv(\tau_P)] \in \mathbb{C}^{N_{\rm s} \times P}$, and also stack $\gv_p$ for all tagets together as $\Gm \overset{\Delta}{=} [\gv_1, \cdots, \gv_P] \in \mathbb{C}^{N_{\rm b}\times P}$, the radar response matrix $\Tm$ can be reformulated as 
%\begin{align}\label{eq:Q_B_G}
    $\Tm = \Bm \Gm^{\rm T}.$
%\end{align}
\vspace{-1mm}
\subsection{Problem Formulation}
%In this subsection, we formulate the considered problem in a concise mathematical model.
Without loss of generality, let us consider the transmission for sensing purpose with $L$ time slots. Define ${\bf{R}} \buildrel \Delta \over = {\left[ {{{\bf r}_1},{{\bf r}_2},\cdots,{{\bf r}_L}} \right]} \in \mathbb{C}^{N_{\rm s} \times L}$ the collection of the observation vectors at all time slots. Then, according to~\eqref{r_l_on-grid},  and $\Tm$, %\eqref{eq:q_b_g}, and \eqref{eq:Q_B_G}, 
the observation matrix for $L$ time slots is given by 
%\begin{align}
%{\bf{R}} &= {\bf{T}} {\bf{W}} + {\bf{Z}} \notag\\
%&= \sum\nolimits_{p = 1}^P {{\bf{b}}\left( {{\tau _p}} \right)} {\bf{g}}_p^{\rm{T}}{\bf{W}} + {\bf{Z}} 
%= {\bf B} {\bf G}^{\rm T} {\bf W} + {\bf Z}. \label{R_mtc_2}
%\end{align}
\begin{align}
{\bf{R}} &= {\bf{T}} {\bf{W}} + {\bf{Z}} 
= \sum\nolimits_{p = 1}^P {{\bf{b}}( {{\tau _p}})} {\bf{g}}_p^{\rm{T}}{\bf{W}} + {\bf{Z}} 
= {\bf B} {\bf G}^{\rm T} {\bf W} + {\bf Z}. \label{R_mtc_2}
\end{align}
where ${\bf W} = [{\bf w}_1, \cdots, {\bf w}_L]$ is the \textit{beam scheduling matrix} of size $N_{\rm b} \times L$, whose $l$-th column is given by ${\bf w}_l$ in \eqref{def:bfw_l}, and ${\bf{Z}} \buildrel \Delta \over = {\left[ {{{\bf z}_1},{{\bf z}_2},\cdots,{{\bf z}_L}} \right]}$ is the effective noise matrix.
% Also note that $\bf Q$ can be decomposed by ${\bf{Q}} \overset{\Delta}{=} \sum\limits_{p = 1}^P {{\bf{b}}\left( {{\tau _p}} \right)} {\bf{g}}_p^{\rm{T}}$, 
% where %${{\bf{b}}\left( {{\tau _p}} \right)}$  given by
 %whose $i$-th element is given by ${h_p}{{\bf{f}}_i^{\rm{H}}{\bf{a}}\left( {{\phi _p}} \right){{\bf{a}}^{\rm{H}}}\left( {{\phi _p}} \right){{\bf{f}}_i}}$. %\nbl{ give the support of $g_p$}.
% Thus, we can rewrite~\eqref{R_mtc} by 
% \begin{align}%\label{R_mtc}
% {\bf{R}} &= \sum\limits_{p = 1}^P\left( {{\bf{b}}\left( {{\tau _p}} \right)}  \otimes {\bf{g}}_p^{\rm{T}}\right){\bf{W}} + {\bf{Z}}\notag\\
% &= \left[ {{\bf{b}}\left( {{\tau _1}} \right) \otimes {\bf{g}}_1^{\rm{T}},{\bf{b}}\left( {{\tau _2}} \right) \otimes {\bf{g}}_2^{\rm{T}},\cdots,{\bf{b}}\left( {{\tau _P}} \right) \otimes {\bf{g}}_P^{\rm{T}}} \right]\left[ \begin{array}{l}
% W\\
% W\\
%  \vdots \\
% W
% \end{array} \right] + {\bf{Z}}.
% \end{align}
While systems of the type of \eqref{R_mtc_2} can be solved for ${\bf G}$ under certain conditions given knowledge of ${\bf B}$ and ${\bf W}$, the lack of knowledge about target delays prevents us from directly applying standard techniques. Nevertheless, we can exploit the structure of the matrices involved and propose an efficient method to estimate the nonzero entries of ${\bf G}$.
\section{The User Acquisition Method}
\label{sec:method}
\vspace{-2mm}
% \nbl{Paper structure: 
% \begin{enumerate}
%     \item First option: elaborate the proposed two methods for user detection, i.e., oversampling delay method, and the two-step method, since we did not come up with a satisfying performance analysis metric. Yi, Shuangyang and I discussed the possibility of gaining some insight of designing the beam scheduling matrix ${\bf W}$. It seems that ${\bf w}$ has to be independently distributed across time slots such that we can use sample covariance matrix to approximate the true covariance matrix for MUSIC algorithm. Or think of a way to approximate the true covariance matrix when the samples are correlated? If ${\bf w}$ is not chosen i.i.d., will the
%     true covariance matrix remains the same for all time slot? If not, should we consider sample covariance matrices of all time slots? Is it possible to analysis the performance (e.g., difference between the true covariance matrix and the sample covariance matrix) when ${\bf w}$ is not chosen i.i.d.
%     \item Second option: focus on the two-step method. Stick to MUSIC algorithm as the  first step. In terms of the second step, we can either employ the traditional MMSE or LASSO algorithm designed for sparsity recovery. The mathematical analysis can be done by assuming the first step is achieved perfectly and determine an upper bound considering different operations of the second step. 
% \end{enumerate}}
%The system model has been well formulated in Section \ref{sec:system}. 
In this section, we introduce a two-stage user acquisition method to address the problem formulated in Section~\ref{sec:system}. Specifically, the proposed method begins by estimating the delay using the MUSIC and subsequently employs a straightforward compressed sensing approach to detect all users in the beam space. Before introducing the details on the proposed methods, we first introduce two beam probing strategies, 
namely, the \textit{beam sweeping strategy} and the \textit{random multi-beam strategy},  focusing on the design of beam weight vectors across different time slots. 
\subsection{Beam Probing Strategies}
\vspace{-2mm}
\label{sec:probing_strategies}
\subsubsection{Beam sweeping strategy}
The beam sweeping strategy aims to prob the signals to all possible beam directions in a sequential manner across different time slots.
Specifically, the beam weight vector at the $l$-th time slot can be expressed as follows,
 %\begin{align}
    $ \wv_{l}^{\rm s
    } = [0, \cdots, 1, 0, \cdots, 0]^{\rm T}, ~\forall l \in [L],$
% \end{align}
 where the only $l$-th element of $\mathbf{w}_l^{\mathrm{s}}$ is set to $1$
 %, while the remaining elements are set to $0$
 , which indicates that we probe in the $l$-th beam direction at the $l$-th time slot. Such a beam probing strategy enables a sufficient power concentration towards the intended beam direction.
\subsubsection{Random multi-beam strategy} 
In this strategy, instead of probing exclusively within a specific beam direction, we employ a random probing approach with $\mathbf{w}_l^{\mathrm{r}}, ~\forall l \in [L]$, and each element of $\wv_l^{\rm r}$ is given by 
%\begin{align}
   $ [\wv_l^{\rm r}]_i = \alpha \widetilde{w}_{l, i}^2, \forall i \in [N_{\rm b}],$
%\end{align}
where $\widetilde{w}_{l, i}$ is unit Gaussian distributed, i.e., $\widetilde{w}_{l, i} \sim \Nc(0, 1)$. Moreover, the power scale factor is set as $\alpha=\frac{1}{N_{\rm b}}$ %should be set as $\frac{1}{N_{\rm b}}$ 
to satisfy the constraint for the beam weight vector that $\mathbb{E} \left[\sum_{i=1}^{N_{\rm b}}[\wv_l^{\rm r}]_i\right] \overset{\Delta}{=}1$. The aim for such a beam probing strategy is to spread the signal power towards many beam directions at the same time.

\vspace{-1mm}
\subsection{ The Two-stage User Acquisition Method}
\vspace{-1mm}
We introduce the two-stage method for user acquisition in the beamspace based on the two proposed probing strategies. %But for notational simplicity, we continue to denote the beam scheduling matrix as $\Wm = [\wv_1, \cdots, \wv_L]$. 
\subsubsection{Stage I: delay estimation using the MUSIC algorithm for $P$ targets}
\label{sec:MUSIC}
Based on the formulation of the received signal in \eqref{R_mtc_2}, assuming that the beam scheduling matrix $\Wm$ and noise $\Zm$ are independent, 
the auto-correlation function of the received signal $\Rm$ is calculated as 
\begin{align}\label{eq:r_cov_form}
    \mathbb{E}[\Rm \Rm^{\rm H}] &= \Bm \Gm^{\rm T} \mathbb{E}[\Wm \Wm^{\rm H}] \Gm^* \Bm^{\rm H} + \mathbb{E}[\Zm \Zm^{\rm H}]\nonumber \\
    &\overset{\Delta}{=} \Bm \Rm_s \Bm^{\rm H} + LN_0 \Id,
\end{align}
where $\Rm_s$ is a matrix with size $P \times P$, defined as 
%\begin{align}
$\Rm_s \overset{\Delta}{=} \Gm^{\rm T} \mathbb{E}[\Wm \Wm^{\rm H}] \Gm^*.$
%\end{align}
In addition, $\Bm \overset{\Delta}{=}[\bv(\tau_1), \cdots, \bv(\tau_P)]$ is a Vandermonde matrix with size $N_{\rm s} \times P$. The form of $ \mathbb{E}[\Rm \Rm^{\rm H}]$ in \eqref{eq:r_cov_form} allows for the use of the MUSIC algorithm to estimate the delays $\tau_p, \forall p \in [P]$ using the eigenspace method.  However, the auto-correlation function of $\Rm$ is not known in advance. It can be estimated by using the samples collected over $L$ times slots,  
%\begin{align}\label{eq:sample_C_r}
    $\widehat{\Cm}_{\Rm} = \Rm \Rm^{\rm H}.$
%\end{align}
Since $\widehat{\Cm}_{\Rm} \in \mathbb{C}^{N_{\rm s} \times N_{\rm s}}$ is a Hermitian matrix, the eigenvalue decomposition on $\widehat{\Cm}_{\Rm}$ is denoted as $\widehat{\Um} \widehat{\Lambdam} \widehat{\Um}^{\rm H}$ with decreasing order on the eigenvalues, i.e., $\widehat{\Lambdam} = {\rm diag}(\widehat{\lambda}_1, \cdots, \widehat{\lambda}_{N_{\rm s}})$ and $\widehat{\lambda}_1 \geq \widehat{\lambda}_2 \cdots \geq \widehat{\lambda}_{N_{\rm s}}$, and all of its eigenvectors $\widehat{\Um} = [\widehat{\uv}_1, \cdots, \widehat{\uv}_{N_{\rm s}}]$ are orthogonal. The eigenvectors corresponding to the $P$ largest eigenvalues span the signal subspace $\Pc_{\rm S}$, while the rest of the eigenvectors span the noise space $\Pc_{\rm N}$. Those two subspaces are orthogonal, i.e., $\Pc_{\rm S}  \perp \Pc_{\rm N}$. By exploiting the orthogonal subspaces, the pseudo-spectrum for the delay $\tau$ is defined as 
\begin{align}\label{eq:MUSIC}
    P_{\rm MU}(\tau) ={\|\bv(\tau)^{\rm H} \widehat{\Um}_{\rm N}\|^2}, ~ \forall~0 \leq \tau \leq T,
\end{align}
where $\widehat{\Um}_{\rm N} \overset{\Delta}{=} [\widehat{\uv}_{P+1}, \cdots, \widehat{\uv}_{N_{\rm s}}]$ denotes the eigenvectors of the noise subspace.
Then the MUSIC algorithm can estimate the delays $\{\widehat{\tau}_1, \cdots, \widehat{\tau}_P\}$ by identifying $P$ dominant minimizers of the pseudo-spectrum $P_{\rm MU}(\tau)$.

% {\color{red} The expression below indicates the delays of all targets are equal to the one that minimizes the pseudospectrum\cdotsI would either just describe informally that we get $P$ peaks and take them as our delay estimates, or somehow define that mathematically, although that is not very immediate.}
% \begin{align}\label{eq:MUSIC}
%     \widehat{\tau}_p = \arg \min_{\tau} P_{\rm MU}(\tau).
% \end{align}

\subsubsection{Stage II: beamspace estimation using the LASSO algorithm}
After applying the MUSIC algorithm on the received signals, the delay estimation for $P$ targets can be obtained as 
%\begin{align}\label{eq:delay_estimate}
    $\widehat{\tauv} = [\widehat{\tau}_1, \cdots, \widehat{\tau}_P].$
%\end{align}
Hence, %based on \eqref{eq:delay_estimate}, 
the delay steering vectors based on delay estimates can be reconstructed as 
%\begin{align}
    $\widehat{\Bm} = [\bv(\widehat{\tau}_1), \cdots, \bv(\widehat{\tau}_P)]$.
%\end{align}
According to \eqref{R_mtc_2}, the received signal in \eqref{R_mtc_2} can be reformulated as $\Rm  \approx \widehat{\Bm} \Gm^{\rm T} \Wm + \Zm,$
%\begin{align*}
%    \Rm  \approx \widehat{\Bm} \Gm^{\rm T} \Wm + \Zm, 
%\end{align*}
which can be rewritten in vectorized form as
\begin{align}\label{eq:vec_R_two_step}
    {\rm vec} ({\bf R}) \approx ({\bf W}^{\rm T} \otimes {\bf \widehat{B}}) {\rm vec}({\bf G}^{\rm T}) + {\rm vec}({\bf Z}),
\end{align}
%The form of \eqref{eq:vec_R_two_step} 
lying in the type of the compressive sensing problem.
Then the beam space information $\Gm$ can be estimated using the LASSO, given as 
%\begin{subequations}
    \begin{align} \label{eq:lasso_two_stage_method}
    \min_{\gv} \quad &\|{\rm vec} ({\bf R}) - ({\bf W}^{\rm T} \otimes {\bf \widehat{B}}) \gv\|^2_2 + \beta\|\gv\|_1,
\end{align}
where $\beta$ is a regularization parameter controlling the sparsity of the solution and we have defined $\gv \buildrel \Delta \over = {\rm vec}({\bf G}^{\rm T})$. In particular, we solve \eqref{eq:lasso_two_stage_method} for different values of $\beta$ and among the $P$-sparse solutions we choose the one which minimizes the reconstruction error.
%\end{subequations}
Let $\widehat{\Gm} \in \mathbb{C}^{P \times N_{\rm b}}$ denote the estimate beamspace matrix from \eqref{eq:lasso_two_stage_method}, and thus for the $p$-th target corresponding to the estimate delay $\widehat{\tau}_p$ from the MUSIC algorithm in \eqref{eq:MUSIC}, the estimate beam index of this target is given by 
%\begin{align}
    $b^{\rm index}_p = \arg \max_{i \in [N_{\rm b}]} |[\widehat{\Gm}]_{p, i}|^2$.
%\end{align}
\vspace{-1mm}
\section{Performance Analysis Using Pairwise Error Probability}
\vspace{-1mm}
\label{sec: performance}
To evaluate the accuracy of the proposed method, we employ the PEP framework to study the theoretical error performance. The PEP quantifies the probability of incorrectly detecting the variable of interest by its distorted version, which provides valuable insights for the design of beam probing strategies. 
% For analytical convenience, we assume that the the beam scheduling matrix $\Wm$, and the delays $\tauv = [\tau_1, \cdots, \tau_P]^{\rm T}$ are available at the receiver.
This analysis also assumes the genie-aided delay estimation, i.e., the delays $\tauv$ are presumed to be known in advance.
According to \eqref{R_mtc_2}, the vectorization of $\Rm$ can be reformulated as  
\begin{align}\label{eq:vec_R_PEP}
    {\rm vec}(\Rm) %= \begin{bmatrix}
%\sum_{p=1}^P h_p \bv(\tau_p) \qv_p^{\rm T} \wv_1\\
%\sum_{p=1}^P h_p \bv(\tau_p) \qv_p^{\rm T} \wv_2\\
%.\\
%.\\
%.\\
%\sum_{p=1}^P h_p \bv(\tau_p) \qv_p^{\rm T} \wv_L
%\end{bmatrix}
%+ 
%\begin{bmatrix}
%\zv_1\\
%\zv_2\\
%.\\
%.\\
%.\\
%\zv_L
%\end{bmatrix} \nonumber \\
%&=\underbrace{\begin{bmatrix}
%\bv(\tau_1) \qv_1^{\rm T} \wv_1 & \cdots & \bv(\tau_P) \qv_P^{\rm T} \wv_1\\
%\bv(\tau_1) \qv_1^{\rm T} \wv_2 & \cdots & \bv(\tau_P) \qv_P^{\rm T} \wv_2\\
%\cdot\\
%\cdot\\
%\cdot\\
%\bv(\tau_1) \qv_1^{\rm T} \wv_L & \cdots & \bv(\tau_P) \qv_P^{\rm T} \wv_L
%\end{bmatrix}}_{(a)}
&=\begin{bmatrix}
\bv(\tau_1) \qv_1^{\rm T} \wv_1 & \cdots & \bv(\tau_P) \qv_P^{\rm T} \wv_1\\
\bv(\tau_1) \qv_1^{\rm T} \wv_2 & \cdots & \bv(\tau_P) \qv_P^{\rm T} \wv_2\\
\cdot\\
\cdot\\
\cdot\\
\bv(\tau_1) \qv_1^{\rm T} \wv_L & \cdots & \bv(\tau_P) \qv_P^{\rm T} \wv_L
\end{bmatrix}v
\!\begin{bmatrix}
h_1\\
h_2\\
\cdot\\
\cdot\\
\cdot\\
h_P
\end{bmatrix}
\!\!+\!\! \begin{bmatrix}
\zv_1\\
\zv_2\\
\cdot\\
\cdot\\
\cdot\\
\zv_L
\end{bmatrix} \nonumber \\
&\overset{\Delta}{=} \Dm_{\tauv, \Wm}(\qv) \hv + \zv,
\end{align}
where we define ${\rm vec}(\Zm)$ as $\zv$, and 
the matrix $\Dm_{\tauv, \Wm}(\qv)$ is a function of the beamspace vector $\qv$ depending on the delays $\tauv$ and the beam scheduling matrix $\Wm$.
For a given channel realization and beam scheduling, let $\widehat{\qv}$ denote the estimate beamspace vector. By noticing that $\hv \sim \Cc\Nc(\mathbf{0}, \Lambdam_{\hv})$, where $\Lambdam_{\hv}$ is a diagonal matrix with its $i$-th diagonal element as $\mathbb{E}[|h_i|^2], ~\forall i \in [P]$,  we shall define the conditional Euclidean distance $d^2_{\hv, \tauv, \Wm} (\qv, \widehat{\qv})$ between the true $\qv$ and estimate beamspace vector $\widehat{\qv}$ as 
\begin{align}
    &d^2_{\hv, \tauv,  \Wm} (\qv, \widehat{\qv}) \overset{\Delta}{=} \|\left(\Dm_{\tauv, \Wm}(\qv) - \Dm_{\tauv,  \Wm}(\widehat{\qv})\right) \hv\|^2 \nonumber \\
    &= \hv^{\rm H} \left(\Dm_{\tauv,  \Wm}(\qv) \!-\! \Dm_{\tauv, \Wm}(\widehat{\qv})\right)^{\rm H} \underbrace{ \left(\Dm_{\tauv,  \Wm}(\qv) \!-\! \Dm_{\tauv,  \Wm}(\widehat{\qv})\right)}_{\widetilde{\Dm}_{\tauv,  \Wm} (\Em)} \hv \nonumber \\
    &=(\underbrace{\Lambdam_{\hv}^{-1/2}\hv}_{\widehat{\hv}})^{\rm H} \underbrace{\left( \Lambdam_{\hv}^{1/2} {\widetilde{\Dm}_{\tauv,  \Wm} (\Em)}^{\rm H} {\widetilde{\Dm}_{\tauv,  \Wm} (\Em)} \Lambdam_{\hv}^{1/2}\right)}_{\Pm_{\tauv, \Wm}(\qv, \widehat{\qv})}  \Lambdam_{\hv}^{-1/2} \hv \nonumber \\
    &= \widehat{\hv}^{\rm H} \Pm_{\tauv, \Wm} (\qv, \widehat{\qv}) \widehat{\hv}
\end{align}
where we define $\widetilde{\Dm}_{\tauv,  \Wm} (\Em) \in \mathbb{C}^{LN_{\rm s} \times P} \overset{\Delta}{=} \left(\Dm_{\tauv,  \Wm}(\qv) - \Dm_{\tauv,  \Wm}(\widehat{\qv})\right)$ corresponding to the detection difference matrix $\Em = [\ev_1, \cdots, \ev_P]$ with $\ev_p = \qv_p - \widehat{\qv}_p, ~\forall p \in [P]$. Let us further define the \textit{squared beamspace difference matrix} $\Pm_{\tauv, \Wm} (\qv, \widehat{\qv}) \overset{\Delta}{=}  \Lambdam_{\hv}^{1/2} \widetilde{\Dm}_{\tauv,  \Wm} (\Em)^{\rm H}\widetilde{\Dm}_{\tauv,  \Wm} (\Em) \Lambdam_{\hv}^{1/2} $, and $\widehat{\hv} \overset{\Delta}{=} \Lambdam_{\hv}^{-1/2}\hv$, which satisfies $\widehat{\hv} \sim \Cc\Nc(\mathbf{0}, \Id_P)$. Since $\zv \sim \Cc\Nc(\mathbf{0}, N_0 \Id)$, the PEP is upper-bounded by~\cite{li2021performance} 
%\begin{align}
    ${\rm Pr}(\qv \rightarrow \widehat{\qv} |\widehat{\hv},  \tauv,  \Wm) \leq \exp\left(-\frac{1}{4N_0} d^2_{\hv, \tauv, \Wm} (\qv, \widehat{\qv})\right).$
%\end{align}
Assuming that the rank of $\Pm_{\tauv, \Wm} (\qv, \widehat{\qv})$ is $r$ and $r \leq \min\{LN_{\rm s}, P\}$, let the eigenvalue decomposition on $\Pm_{\tauv, \Wm} (\qv, \widehat{\qv})$ define as $\Um\Lambdam {\Um}^{\rm H}$, where $\Um= [\uv_{1}, \cdots, \uv_{P}]$ denotes the eigenvectors and the corresponding eigenvalues are denoted as the diagonal elements of $\Lambdam$, i.e., $\Lambdam = {\rm diag}\left(\lambda_1, \cdots, \lambda_P\right)$. Note that all the eigenvalues are non-negative since $\Pm_{\tauv, \Wm} (\qv, \widehat{\qv})$ is a positive semi-definite matrix. Therefore, the PEP can be reformulated as 
\begin{align}\label{eq: pep_tilde_h}
   {\rm Pr}(\qv \rightarrow \widehat{\qv} |\widehat{\hv},  \tauv,  \Wm) &\leq \exp\bigg(-\frac{1}{4N_0} \widehat{\hv}^{\rm H}\big(\sum_{p=1}^r \lambda_p \uv_p {\uv_p^{\rm H}} \big) \widehat{\hv}\bigg) \nonumber \\
   %&=\exp\left(-\frac{1}{4N_0} \sum_{p=1}^r \lambda_p \widehat{\hv}^{\rm H}\uv_p \uv_p^{\rm H}\widehat{\hv} \right) \nonumber \\
   &= \exp\bigg(-\frac{1}{4N_0} \sum_{p=1}^r \lambda_p|\widetilde{h}_p|^2\bigg),
   \vspace{-0.1cm}
\end{align}
where  $\widetilde{h}_p \overset{\Delta}{=} \uv_p^{\rm H} \widehat{\hv}, ~\forall p \in [P]$.
% Since the eigenvectors $\Um^{\Pm} = [\uv_1^{\Pm}, \cdots, \uv_P^{\Pm}]$ depend on the delays $\tauv$, and channel gains $\hv$ also depend on the delays, $\Um$ and $\hv$ are independent conditioned on the delays. 
It can be shown that $\{\widetilde{h}_1, \cdots, \widetilde{h}_P\}$ are independent complex Gaussian random variables (RVs) with mean $\mu_{\widetilde{h}_p} = \uv_p^{\rm H} \mathbb{E}[\hv] = 0$ and variance $C_{\widetilde{h}_p} = \uv_p^{\rm H} \mathbb{E}[\hv \hv^{\rm H}] \uv_p = 1$.
Thus, it is obvious that $|\widetilde{h}_p|$ follows the Rician distribution with a Rician factor $K_p = |\mu_{\widetilde{h}_p}|^2, ~\forall p \in [P]$, and its probability density function (pdf) is given by 
%\begin{align}
    $p(|\widetilde{h}_p|) = |\widetilde{h}_p| \exp(-\frac{|\widetilde{h}_p|^2}{2} - \frac{K_p}{2}) I_0(\sqrt{K_p}|\widetilde{h}_p| ),$
%\end{align}
where $I_0(\cdot)$ is the modified Bessel function and $I_0(0) = 1$. Herein, $K_p = 0$, then the pdf can be simplified as 
%\begin{align}
    $p(|\widetilde{h}_p|) = |\widetilde{h}_p| \exp(-\frac{|\widetilde{h}_p|^2}{2}).$
%\end{align}
Note that $\widehat{\hv}$, and $\tauv$ are independent from each other. Hence, the conditional PEP dependent only on $\tauv$ and $\Wm$ can be obtained based on the fact that ${\rm Pr}(\qv \rightarrow \widehat{\qv} |\tauv,  \Wm) = \int {\rm Pr}(\qv \rightarrow \widehat{\qv} |\widehat{\hv},  \tauv,  \Wm) p(\widehat{\hv}) d \widehat{\hv}$. Since $\{\widetilde{h}_1, \cdots, \widetilde{h}_P\}$ are independent complex Gaussian RVs, the averaging over $|\widetilde{h}_p|$ can be derived as 
\begin{align}\label{eq: int_h_p}
    &\int_{0}^{\infty} \exp\Big(-\frac{1}{4N_0}  \lambda_p|\widetilde{h}_p|^2\Big) p(|\widetilde{h}_p|) d|\widetilde{h}_p| \nonumber \\
    =&\int_{0}^{\infty} \exp\Big(-\frac{1}{4N_0}  \lambda_p|\widetilde{h}_p|^2\Big) |\widetilde{h}_p| \exp(-\frac{|\widetilde{h}_p|^2}{2}) d|\widetilde{h}_p|  \nonumber \\
    =& \frac{1}{\frac{1}{2N_0}\lambda_p + 1}.
\end{align}
Therefore, based on \eqref{eq: int_h_p}, ${\rm Pr}(\qv \rightarrow \widehat{\qv} |\tauv,  \Wm)$ is given by 
\begin{align}
    {\rm Pr}(\qv \rightarrow \widehat{\qv} |\tauv,  \Wm) &\leq \prod_{p=1}^r \frac{1}{\frac{1}{2N_0}\lambda_p + 1} 
    \leq \prod_{p=1}^r \left(\frac{1}{2N_0}\lambda_p\right)^{-1} \nonumber\\
    % &= \left(\frac{1}{2N_0}\right)^{-r} \prod_{p=1}^r \left(\lambda_p^{\Pm}\right)^{-1} \nonumber \\
    &=  \Big(\frac{1}{2N_0}\Big)^{-r}\Big( \Big(\prod_{p=1}^r \lambda_p\Big)^{\frac{1}{r}}\Big)^{-r}. \label{eq:conditional_PEP}
    % &=\textcolor{blue}{\frac{1}{\det(\Pm_{\tauv, \Wm} (\qv, \widehat{\qv}))}\left(\frac{1}{2N_0}\right)^{-r}}
\end{align}
From~\eqref{eq:conditional_PEP}, it is evident that the performance of the considered problem depends on the rank and the geometric mean of non-zero eigenvalues, i.e., $\tiny{(\prod_{p=1}^r \lambda_p)^{\frac{1}{r}}}$, of the squared beamspace difference matrix.
Particularly, the rank of $\Pm_{\tauv, \Wm} (\qv, \widehat{\qv})$ determines the error exponent of the PEP (commonly referred to as the diversity gain), while its non-zero eigenvalues' geometric mean determines the SNR improvement (commonly referred to as the coding gain). These understandings align well with the conventional PEP analysis for space-time codes~\cite{Calderbank1998PEP}. Therefore, a preliminary guideline for the design of the beam scheduling matrix $\Wm$ in this problem is to maximize both the rank and the geometric mean of non-zero eigenvalues.
To access user acquisition performance, we adopt two standard metrics: missed detection rate and false alarm rate. The missed detection rate $P_{\rm MD}$ is the ratio between the number of missed targets in the beamspace (i.e., the sum of targets that are not detected at the correct beam direction) and number of total targets. The false alarm rate $P_{\rm FA}$ is the ratio between the number of mistakenly detected targets (i.e., the sum of number of detected targets minus the number of actual targets at each beam direction) and the number of total targets.  %whose definitions are given as follows.

\section{Simulation Results}
\label{sec: simulation}
In our simulation, we consider $M= 128$ antennas, $P=3$ targets, bandwidth ${\rm BW} = 160$MHz, the carrier frequency $f_c = 10$GHz, $N_{\rm s} =36$ subcarriers, the noise power spectral density $N_0=-174$dBm/Hz, and the number of beam directions $N_{\rm b} = 36$. In addition, for each target $p$, we consider the relative distance of each target is uniformly distributed between $10$m and $50$m, i.e., $d_p \sim {\rm Unif}[10, 50]$, the radar cross section $\sigma_{p, {\rm rcs}} = 20$ dBsm, and we also assume that the AoA is uniformly distributed between $-\frac{\pi}{2}$ and $\frac{\pi}{2}$. The numerical results are averaged over $50$ channel
geometries and each with $100$ channel realizations. To shed the light on the system design, we will provide quantitative comparisons of the two beam probing strategies provided in Section III.

\subsection{Evaluation for the Squared Beamspace Difference Matrix}
\label{sec:sim_comparison}
% \begin{figure}[ht!]
% %\centering
%         %\begin{subfigure}[b]{0.42\textwidth}
%         \centerline{ \includegraphics[width=0.45\linewidth,trim={19 34 18 19}]%\columnwidth]
%         {figures/rank.pdf} \hspace{.5cm}
%         %\caption{\blue{The probability distribution of the ranks of $\Pm_{\tauv, \Wm}(\qv, \widehat{\qv})$ with the two beam probing strategies at different time slots $L = \{6, 18, 36\}$.}}
%         %\label{fig:diversity_gain}
%         %\end{subfigure}
%         %~~
%         %\begin{subfigure}[b]{0.4\textwidth}
%     \includegraphics[width=0.45\linewidth, trim={19 34 18 19}]%[width=\columnwidth]
%     {figures/determinant.pdf} 
%         }
%        % \caption{\blue{The cumulative distribution function of $\left(\prod_{p=1}^r \lambda_p\right)^{\frac{1}{r}}$ with two beam probing strategies at different time slots $L = \{6, 18, 36\}$.}}
%        % \label{fig:coding_gain}
%         %\end{subfigure}
%         \vspace{.5cm}
%  \caption{ Left:. Right: .  %{\color{blue} SL: Diversity and coding gain of two beam probing strategies.} 
%  }
%       \vspace{-4mm}
% \end{figure}

\begin{figure}[t!]
\centerline{ \includegraphics[width=0.49\linewidth]{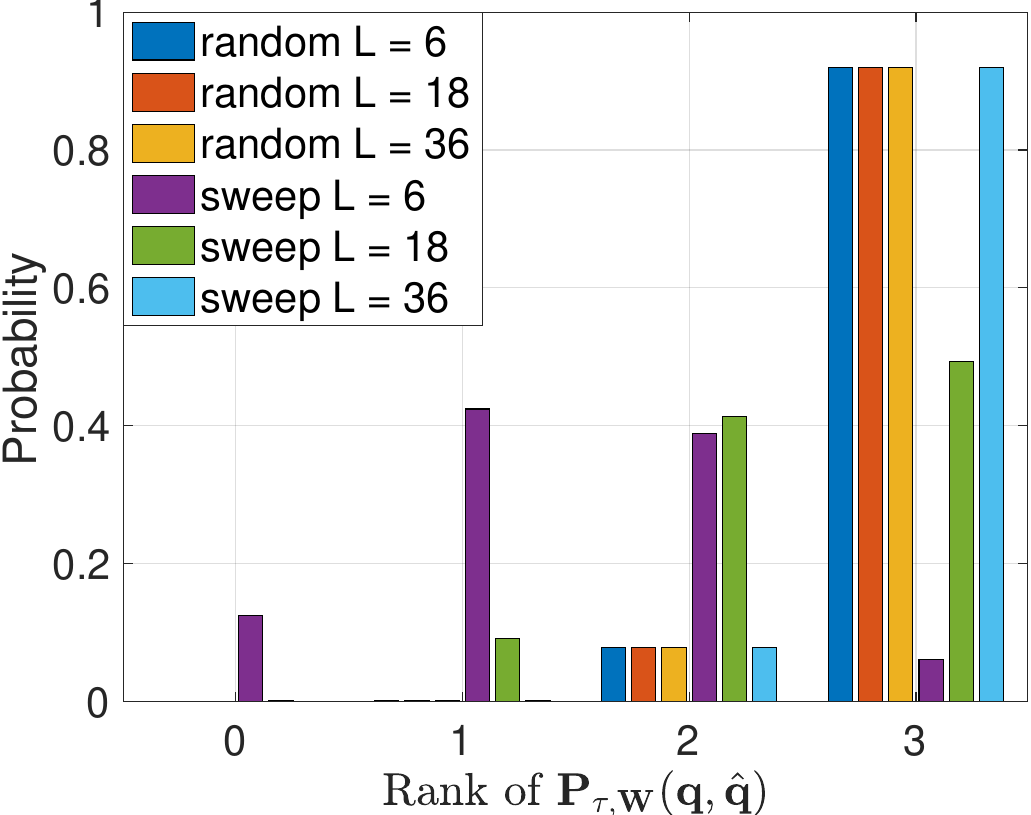} \hspace{.0cm} 
\includegraphics[width=0.49\linewidth]{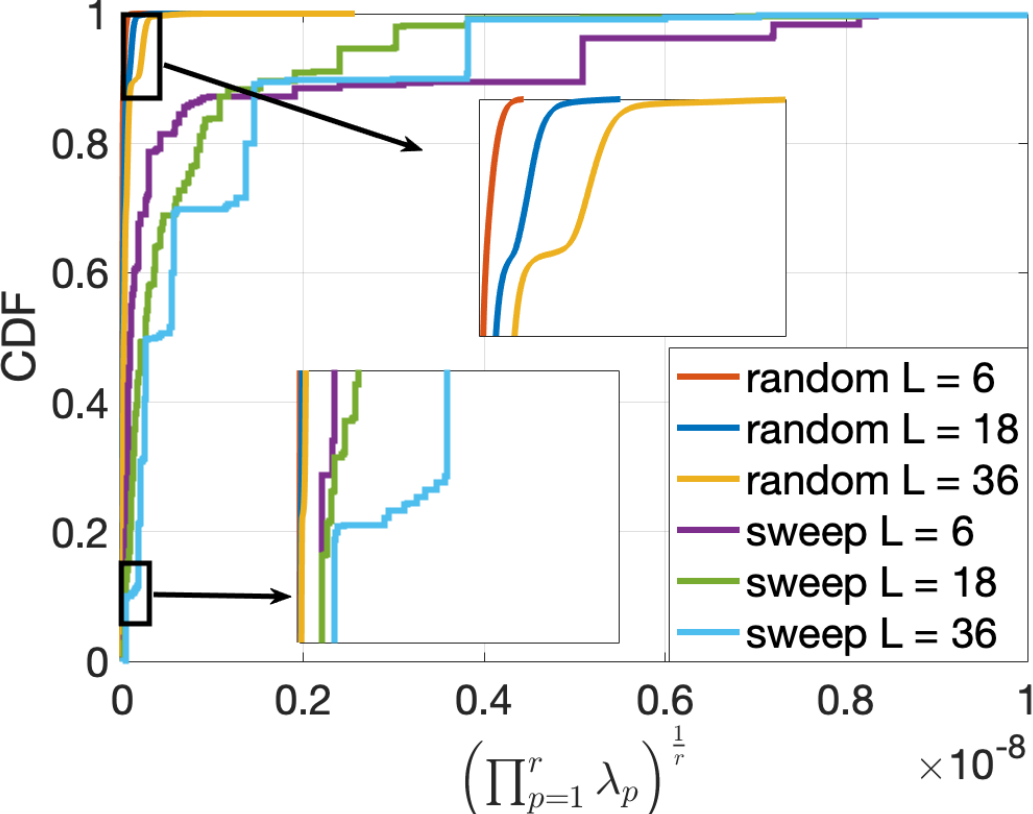} }
\caption{Left:Evaluation on ranks. Right:Evaluation on geometric means.}
        %\begin{subfigure}[b]{0.35\textwidth}
        %\includegraphics[width=\columnwidth]{figures/rank.pdf}
        %\caption{Evaluation on ranks.}
        %\label{fig:diversity_gain}
        %\end{subfigure}
       %~~~~~~~~~~~~~~~~
        %\begin{subfigure}[b]{0.34\textwidth}
        %\includegraphics[width=\columnwidth]{figures/determinant.pdf}
        %\caption{Evaluation on geometric means.}
        %\label{fig:coding_gain}
        %\end{subfigure}
 %\caption{Evaluation on the beamspace matrix in terms of matrix rank and geometric mean of non-zero eigenvalues.}
     \label{fig:probing_comparison}
	\vspace{-.55cm}
\end{figure}
Based on the analysis in Section \ref{sec: performance}, we conduct a comparative evaluation of the two proposed beam probing strategies in terms of the rank and the geometric mean of non-zero eigenvalues for $L=\{6, 18, 36\}$ in Fig.~\ref{fig:probing_comparison}. The evaluation is conducted over 50 instances of true beamspace vectors $\qv$ and all the possible distorted beamspace vectors. As shown in Fig.~\ref{fig:probing_comparison}, %{fig:diversity_gain}, 
we observe that the probability distribution of ranks for the random multi-beam strategy remains relatively consistent across different values of $L$. On the other hand, for the beam sweeping strategy, the probability on the high rank increases as time slots increase. This behavior is attributed to
the fact that the beam sweep strategy is unlikely to prob the signal to the desired direction within a short period of time.
Consequently, $(\qv_i - \widehat{\qv}_i)^{\rm T} \wv_l^{\rm s}, ~\forall l \in [L]$, are all equal to zero, leading to a decrease in the rank of the squared beamspace difference matrix. 
Furthermore, we notice that the ranks corresponding to the two probing strategies converge to the same value when $L=36$. The above observations suggest that
the random multi-beam strategy enables a quick decay of the error performance compared to the beam sweeping strategy when the available time slots are less. However, when the system operates with sufficient time slots, the slopes of error curves of these two strategy are roughly the same. 

We demonstrate the geometric mean of non-zero eigenvalues of the squared beamspace difference matrices corresponding to the two beam probing strategies in
Fig.~\ref{fig:probing_comparison}, %{fig:coding_gain}, 
where the beam sweeping strategy generally enjoys a larger value compared to the random multi-beam counterpart. Notice that the improvement of geometric mean of non-zero eigenvalues corresponds to the potential SNR improvement when the number of time slots available is sufficiently large. This observation indicates a superior error performance shall be observed for the beam sweeping strategy given a sufficient number of time slots, e.g., $L \geq 36$.

%This suggests that the PEP decreases more rapidly when the random multi-beam strategy is applied. However, when the ranks for the two proposed probing strategies at $L=36$ are the same, the PEP is mainly affected by the SNR improvement corresponding to the geometric mean of non-zero eigenvalues. As illustrated in Fig. \ref{fig:coding_gain}, when we compare the two probing strategies at $L=36$, it is evident that the potential SNR improvement\footnote{Although the eigenvalues are quite small in the simulations, they make sense when taking the super low noise power spectral density $N_0$ in the radar system into account.} of the beam sweep strategy is larger than that of the random multi-beam strategy with high probability, which implies the good performance of the beam sweep strategy given enough time slots, e.g., $L=36$.

% \subsection{The effect of the probing strategies on the MUSIC algorithm}
% Next, we investigate the performance of the MUSIC algorithm under different beam probing strategies. 
\subsection{Missed detection rate and false alarm rate vs. time slots}
\begin{figure*}[t!]
\centerline{\includegraphics[width=0.49\linewidth]{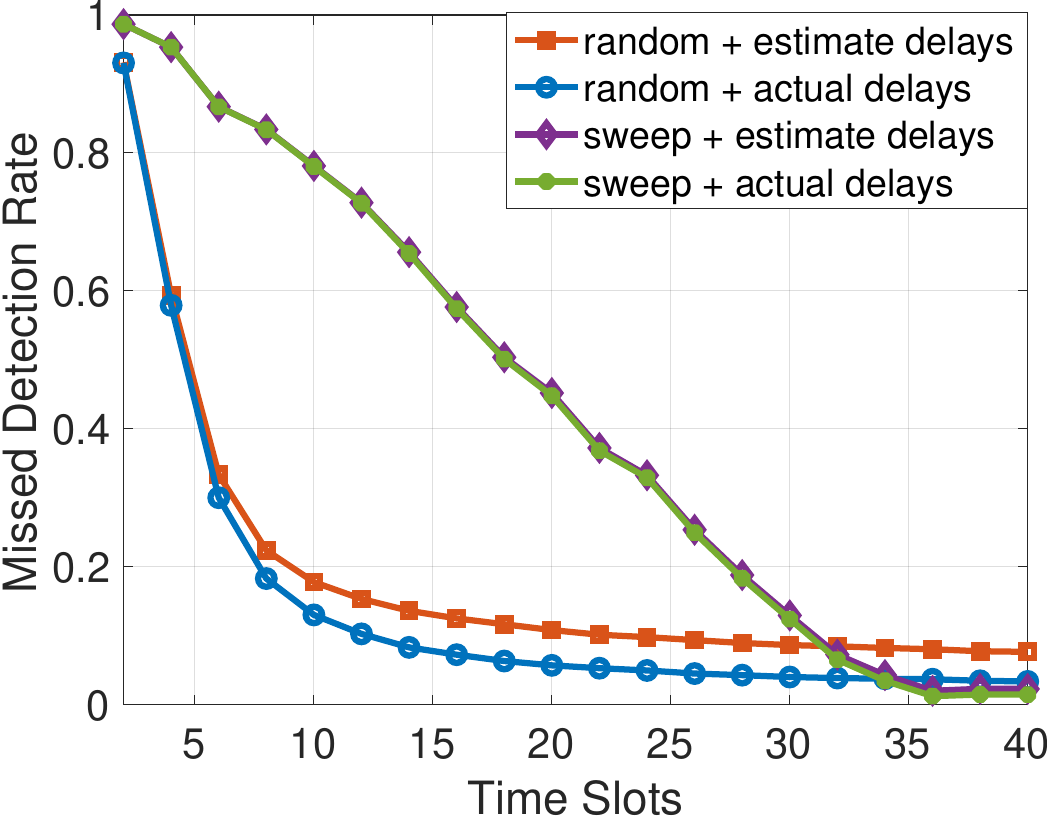} \hspace{.0cm}
\includegraphics[width=0.49\linewidth]{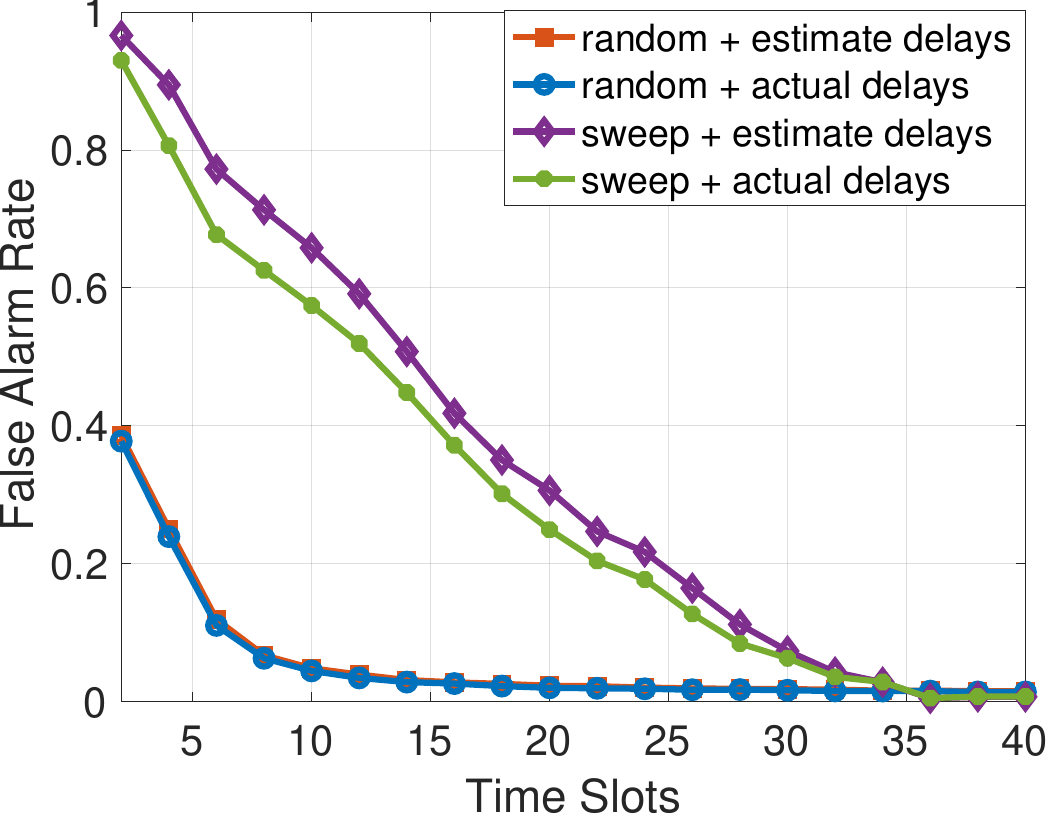} }
        %\begin{subfigure}[b]{0.33\textwidth}
        %\includegraphics[width=\columnwidth]{figures/MD.pdf}
        \caption{Left: Evaluation on missed detection. Right: Evaluation on false alarm.}
        %\label{fig: MD}
        %\end{subfigure}
        %~~~~~~~~~~~~~~~~
        %\begin{subfigure}[b]{0.33\textwidth}
        %\includegraphics[width=\columnwidth]{figures/FA.pdf}
        %\caption{Evaluation on false alarm}
        \label{fig: FA}
        %\end{subfigure}
 %\caption{The user acquisition metrics of the two-stage method under two probing strategies with and without the explicit information on the actual delay.}
        \label{fig:MD_FA_L}
\end{figure*}

% \begin{figure}[ht!]
% %\centering
%         %\begin{subfigure}[b]{0.42\textwidth}
%         \centerline{ \includegraphics[width=0.45\linewidth,trim={19 34 18 19}]%\columnwidth]
%         {figures/MD.pdf} \hspace{.5cm}
%         %\caption{\blue{The probability distribution of the ranks of $\Pm_{\tauv, \Wm}(\qv, \widehat{\qv})$ with the two beam probing strategies at different time slots $L = \{6, 18, 36\}$.}}
%         %\label{fig:diversity_gain}
%         %\end{subfigure}
%         %~~
%         %\begin{subfigure}[b]{0.4\textwidth}
%     \includegraphics[width=0.45\linewidth, trim={19 34 18 19}]%[width=\columnwidth]
%     {figures/FA.pdf} 
%         }
%         \vspace{.5cm}
%   \caption{}
  
        %\vspace{-6mm}
% \end{figure}       
\iffalse
\begin{figure*}[ht!]
\centering
        \begin{subfigure}[b]{0.4\textwidth}
        \includegraphics[width=\columnwidth]{figures/MD.pdf}
        \label{fig:MD_L}
        % \caption{$P_{\rm MD}$ vs. $L$.}
        \end{subfigure}
        %~~
        \begin{subfigure}[b]{0.4\textwidth}
        \includegraphics[width=\columnwidth]{figures/FA.pdf}
        \label{fig: FA_L}
        % \caption{$P_{\rm FA}$ vs. $L$.}
        \end{subfigure}

 \caption{The comparisons for the two-stage method under different probing strategies and delay information in terms of $P_{\rm MD}$ and $P_{\rm FA}$ vs. time slots $L$.}
     \label{fig:MD_FA_L}
\end{figure*}
\fi 
In Fig. \ref{fig:MD_FA_L}, we show the comparison between the two beam probing strategies using the proposed two-stage method. As a performance benchmark, we also plot the error performance of the LASSO by assuming the actual delay of the users are known in prior, which serves as a lower bound for the proposed two-stage method. %In Fig.~\ref{fig: MD}, 
We observe that the random multi-beam strategy enjoys a better missed detection rate compared to the beam sweeping strategy with less number of time slots. However, the beam sweeping strategy outperforms the random multi-beam strategy when the number of time slots is sufficiently large, e.g., $L \ge 36$. These observations agree with the insights we obtained from the evaluations of the squared beamspace difference matrix. Furthermore, we also notice that the two-stage method performs general well, whose performance approaches to the results with known actual delay.
We observe similar results from the false alarm rate results in Fig.~\ref{fig: FA}. Specifically, we observe that the random multi-beam strategy enjoys a superior performance with less time slots but suffers performance loss with sufficiently large time slots compared to the beam sweeping strategy. 
Based on the above discussions, we conclude that the random multi-beam strategy is favorable when the size of the beamspace matrix is larger than the available time slots. On the other hand, the beam sweeping strategy is a better choice for beamspace matrix  with a relatively small size.

\section{Conclusion}
In this work, we focused on radar-assisted user acquisition for downlink transmissions in multi-user MIMO OFDM systems. We first proposed a two-stage method that includes the initial estimation of the delays using the MUSIC algorithm, followed by the user acquisition through the estimation of the beam space responses using a compressed sensing method. 
Furthermore, we conducted theoretical performance analysis based on the pairwise error probability framework, which unveils important design criteria for beam probing. Our numerical results align with our analysis and verify the effectiveness of the proposed two-stage method.

%We also consider the two beam probing strategies, i.e., the beam sweep and the random multi-beam strategy, and derive some theoretical analysis based on the PEP, which provides the theoretical insights on the comparison between the two strategies. The numerical results strongly emphasize the advantageous performance of the random multi-beam strategy, especially in scenarios with limited time slots, which is also consistent with our theoretical findings.
%{\color{blue}SL: In this work, we focused on the downlink integrated sensing and communication scenario, and formulated the user detection problem from the compressed sensing perspective. }

\bibliographystyle{IEEEtran}
\bibliography{ref}

\end{document}